\documentclass{article}
\usepackage{sao2,psfig}
\issue{2002, 54,  }

\begin{document}
\newcommand{\km}{\,\mbox{km} \cdot \mbox{s}^{-1}}
\markboth{Moiseev}{Reduction of CCD observations with scanning Fabry-Perot
interferometer}
\setcounter{page}{1}
\title{Reduction of CCD observations with scanning Fabry-Perot interferometer}
\author{A.V. Moiseev}
\institute{\saoname}
\date{March 18, 2002}{July 11, 2002}
\maketitle

\begin{abstract}
The algorithms used for reduction of data obtained in a Fabry-Perot
interferometer mode of the new focal reducer of the 6m telescope SCORPIO are
described. The main attention is focused on the procedures of photometry
correction of channels and night sky emission subtraction, which is especially
important for observations with a CCD in comparison with a photon counter. The
influence of various sources of errors on measurements of radial velocities in
a data cube is investigated. Examples of velocity field construction by using
emission lines in the galaxy NGC\,6951 and absorption lines in the globular
cluster M\,71 are presented. 

\end{abstract}

\section{Introduction}
The scanning Fabry-Perot interferometer (IFP) is a highly efficient device
for investigation of kinematics of extended objects by the method of
two-dimensional spectroscopy. Observations with the IFP consist of successive
acquisition of a few dozen images of interference rings from the object
(or from a calibration lamp) when changing the optical path between the
flat-parallel plates. The radius of the rings is a function of wavelength and
gap between the plates of the interferometer. The complete set of such
images which fills the free spectral range of the interferometer is called
a scanning cycle. After a special reduction, these images may be represented
 as a data cube. In this cube, X and Y (``spatial coordinates'') correspond
to the coordinates in the sky plane, while the wavelengths (or Doppler
velocities in a fixed spectral line) are a third ``spectral coordinate'' Z
(see Fig. 1). For observations of galaxies, such a technique was first applied
by Tully (1974) in studying the motions of ionized gas in M51 in the
$H_{\alpha}$ line. A photographic plate was used as the detector. In the
following two decades the systems with the scanning IFP were employed in
observations at many big telescopes (TAURUS-2  at the 4.2 m telescope WHT
(Spain), PUMA at the 2.1 m telescope of OAN (Mexico), CIGALE at the 3.6m
telescope (ESO), the project KTS for the Japanese 8.2 m telescope SUBARY,
etc.). The study of two-dimensional kinematics of galaxies at the 6m telescope
of SAO RAS with the use of the scanning IFP was started by a group of
researchers from Marseille Observatoire in cooperation with the colleagues
from SAO in the first half of 1980s (Boulesteix et al., 1982; Amram et al.,
1992; Dodonov et al., 1995). The observations were made with the system CIGAL
which consisted of the focal reducer with the IFP and the two-dimensional
photon counter IPCS as the detector. The IPCS was replaced by a low read-out
noise CCD in 1997, and in the September   2000  the first observations with a
new multimode optical reducer SCORPIO (Spectral Camera with Optical Reducer for
Photometrical and Interferometrical Observations) are carried out. As compared
with the old reducer, the SCORPIO is fully automated, its transmission is
several times as high and the optics for accurate calibration is better.

In the present paper we describe the procedure adopted for the reduction of
observations with the SCORPIO in the IFP mode. At the present time, there
are available several extensively used programmes of reduction of this kind
of data. First of all, this is the package ADHOS developed by Boulesteix (2000)
and the package of programmes operated under the IRAF (see Gordon et al., 2000).
However, in our opinion, in the programmes mentioned the specific character of
observations with a CCD in which the contribution of the atmosphere in the
data cube varies from frame to frame is not adequately taken into account.
This leads to the appearance of systematic errors and artifacts in the
construction of radial velocity fields. Note that this problem is obviated
in observations with the photon counter which makes short exposures and
thereby eliminates the effect of the atmosphere. The new procedure of reduction
of CCD images that we developed takes account of the effect of the atmosphere
and improves the radial velocity measurement accuracy.

The basic data from the theory of the interferometer are presented in Section
2.1. The present-day facilities used for IFP observations of SAO and the
problems that present themselves with the application of CCDs are discussed in
Sections 2.2 and 2.3. The primary reduction of observations and the conversion
of the data cube to the wavelength scale are described in Sections 3 and 4.
The sources of errors affecting the radial velocity measurement accuracy are
explored in Section 5. The final Section 6 gives results of observations of
two objects.

\begin{figure}
\centerline{\psfig{figure=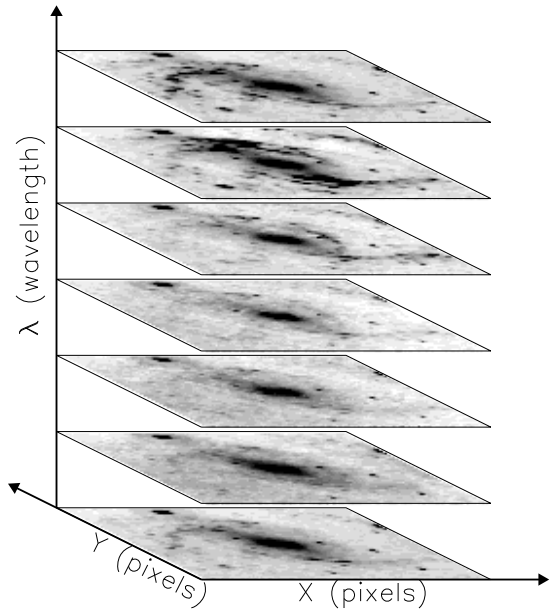,width=7cm}}
\caption{An example of the ``data cube'' obtained with the IFP. Every fourth
channel along the wavelength $\lambda$ is shown.}
\label{cube}
\end{figure}

\section{The Fabry-Perot interferometer of SAO RAS}
\subsection{Principal relations}
The theory of the Fabry-Perot interferometer as applied to astronomical
investigations is stated in detail in a number of papers (Courtes, 1960; Bland
and Tully, 1989; Gordon et al., 2000). Below we will present but a few
necessary relations. If a parallel monochromatic beam of light with a wavelength
$\lambda$ strikes the interferometer at an angle $\vartheta$ with respect
to the optical axis, the condition of maximum of the interference pattern is
then written as
\begin{equation}
n\lambda=2l\mu\cos\vartheta=\frac{2l\mu}{\sqrt{1+\left(\frac{r}{f}\right)^2}}.
 \label{main}
\end{equation}

Here $n$ is the order of interference, $l$ is the distance between the interferometer
plates, $\mu$ is the refractive index of the medium between the plates.
The optical system (a camera with a focal length $f$) constructs the image of
the interference rings on a two-dimensional detector (Fig. 2). Differentiating
(1), we obtain an expression for the angular dispersion:
\begin{equation}
\frac{d\lambda}{d\vartheta}=-\frac{2l\mu}{n}\sin\vartheta=
-\frac{2l\mu}{n}\frac{1}{\sqrt{1+\left(\frac{f}{r}\right)^2}}.
 \label{disp}
\end{equation}

It follows from (2) that, firstly, the dispersion grows towards the centre
of the rings, and, secondly, the width of the rings decreases with increasing
radius $r$. The distance between the adjacent orders of interference defines
the free spectral range of the IFP:
\begin{equation}
\Delta\lambda=\lambda/n.
 \label{resol}
\end{equation}

An important characteristic property of the IFP is its finesse:

\begin{equation}
F=\Delta\lambda/\delta\lambda,
 \label{fin}
\end{equation}
where $\delta\lambda$ is the $FWHM$ of its instrumental profile which defines
the spectral resolution and depends, first of all, on the characteristics of
the reflective coatings of the IFP.

For the scanning interferometer $2\mu l=A+Bz$, where $z$ is the number of
the spectral channel, $A$ and $B$ are certain constants (Gordon et al., 2000).
One can easily find that
\begin{equation}
B=\frac{A}{n\,n_z},
 \label{AB}
\end{equation}
where $n_z$ is the number of channels in the scanning cycle.

\begin{figure}
\centerline{\psfig{figure=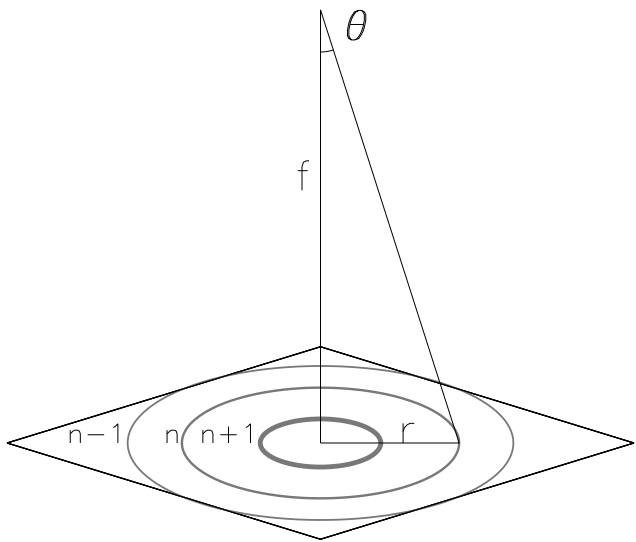,width=8cm}}
 \caption{Interference pattern from monochromatic light. Different rings
correspond to different interference orders  (n-1, n, n+1).}
 \label{theori}
\end{figure}

Only the regions for which condition (1) is satisfied will be seen in the
image of an extended object in channel $z$.  Thus, spatial and spectral
information is mixed in each frame, so that a certain wavelength $\lambda$
corresponds to the point (x,y). From (1) taking into account  that $r\ll f $
and according to (5) $A\gg B$, we obtain an expression for the number of
the channel in which the interference maximum is observed at the given radius:
\begin{equation}
z(r)\cong\frac{n}{B}\lambda + \frac{A}{2Bf^2}r^2-\frac{A}{B}.
 \label{parabola}
\end{equation}

The distribution of the quantity $z(x,y)$ in the image plane for the calibration
lamp line is generally called the phase shift map, and the procedure of transition
from $z$ to $\lambda$ is the phase correction. An example of such a map is displayed in Fig. 3.
Expression (6) shows that the phase shift in a first approximation is
proportional to the square of the distance from the centre of the rings.

\subsection{The IFP mode with the SCORPIO}

The new multimode optical reducer SCORPIO was made at SAO in 2000 (Afanasiev
et al., 2002). We will briefly describe here the potentialities of the SCORPIO
in interferometric observations. The basic optical components of the SCORPIO
are: a collimator $F/2.2$ and a camera $P/1.8$, the total optical efficiency
of the system at the prime focus of the 6m telescope is $F/2.9$. The optics of
the reducer compensate for the aberrations of the main mirror of the
telescope, all optical surfaces have antireflecting coating in a range of
$0.35-1.0\mu$. There are two filter wheels, one is in the focal plane of the
telescope, the other is between the field lens and collimator. When observing
with the IFP, the desired spectral range is cut out with the aid of
interference filters with a $FWHM\approx 10\div 15$\, \AA\, placed in the
filter wheel near the focal plane. Presently, a set of filters with a maximum
transmission of $60-80$\% is used, which are centered
 on the $H_{\alpha}$
line in galaxies with systemic velocities from $-200$ to $\rm +10000\,km\cdot s^{-1}$.
A similar set is manufactured for observations in the line [OIII]$\lambda 5007$.

A piezoelectric IFP ET-50 produced by the company Queensgate is being
installed between the collimator and camera, where the exit pupil of the
optical system is located. The parameters of the  interferometers are
presented in Table 1. In 2000--2001 a CCD system Tecktronix ${\rm 1K\times 1K}$
with a readout noise of $3\bar{e}$ was used in observations. This detector
provided a field of view of about $5\arcmin$ at a scale of $0.28\arcsec$/pixel.
A new CCD of 2K$\times$2K is contemplated to be installed in 2002. The total
quantum efficiency of the SCORPIO in observations with the interferometer
Fabry-Perot (telescope + filter + IFP + CCD) is about 20\,\% in the
$H_{\alpha}$ region.
\begin{figure*}
\psfig{figure=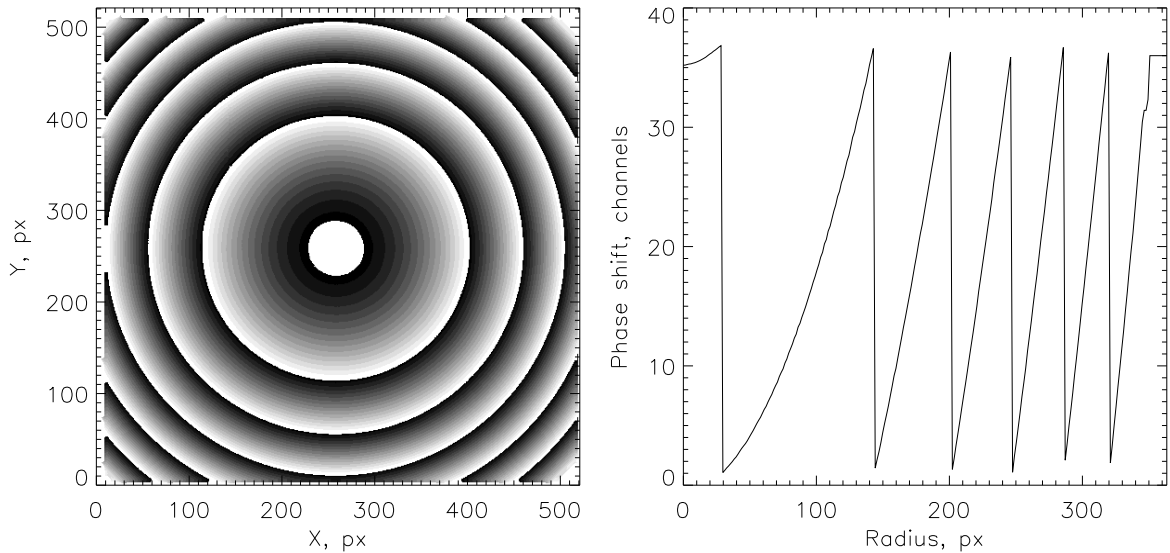,width=16cm}
 \caption{Map of phase shifts for  FP501 (left panel).
Radial variation of the phase shift in units of spectral channel (right panel).}
 \label{phas}
\end{figure*}

\begin{table}
\caption{SCORPIO parameters in IFP mode}
\begin{center}
\begin{tabular}{lll}
\hline
 &\multicolumn{2}{c}{Queensgate IFP}\\
     & FP260 & FP500  \\
\hline
range   &$5000\div7000$\,\AA\, &$5000\div8000$\,\AA\, \\
 $n^a$  & 235 & 501\\
  $\delta\lambda^a$   &$2.5$\,\AA& $0.7$\,\AA\, \\
$\Delta\lambda^a$  & 29\,\AA& 13\,\AA\, \\
 $F^a$ &11  &17 \\
 $n_z$   &$24\div32$& $32\div40$ \\
\hline
%\end{tabular}
%\end{center}
\multicolumn{3}{c}{\footnotesize $^a$ values are given for the wavelength  $6563$\,\AA}
\end{tabular}
 \end{center}
\end{table}

The optical reducer is mounted on the universal adapter of the prime focus,
which contains two movable bundles of fibers for the off-axis guiding, lamps
for the calibration of the wavelength scale (a He-Ne-Ar lamp) and for flat
fielding (a continuous spectrum lamp). The lamps are placed in an integrating
sphere providing uniform illumination of the field. The calibration beam forms
at the input of the reducer a beam of an aperture ratio of $F/4$ equivalent to
that of the telescope. Such a scheme of illumination with equal filling of the
output pupil in observations and calibration ensures the absence of shift when
measuring radial velocities. This was confirmed by real observations with the
IFP on the 6m telescope (see Sections 5.1 and 6.2).

\subsection{Comparison of the photon counter and CCD}
In observations with the photon counter one executes several dozen cycles of
scanning with a very short exposure (10--20 s) in each channel of the IFP. In
this fashion one can manage to average the influence of the atmospheric
extinction at different zenith distances and the instabilities of seeing with
long total exposures. The CCD chip has a considerably higher quantum efficiency
as compared to the photon counter. However, the presence of the readout noise
does not permit very short acquisitions, and the readout time diminishes the
useful exposure time. The use of low-noise CCDs makes it possible to slightly
improve the situation. Employing relationship (3) from the paper by Bland and
Tully (1989), we compared the capabilities of the SCORPIO in reaching a
maximum possible signal-to-noise ratio, when using our ``thin'' CCD TK\,1024
(the readout noise is $3\bar e$, the maximum quantum efficiency is $\sim
80$\,\% at $\lambda 6500$\,\AA) and using a present-day photon counter with a
GaAs cathode and maximum quantum efficiency $\sim 25$\,\% (Hernandez et al.,
2001). The calculations were performed for the observations at the prime focus
of the 6m telescope with the system SCORPIO. In a wide range of intensities of
objects, the CCD reaches a larger S/N ratio only if the acquisition time
exceeds the readout time (($T_{exp}>T_{readout}$). However, if $T_{exp}\simeq
T_{readout}$, the observing time is then spent inefficiently because half of
it will be spent for readout of the signal from the CCD. Now $T_{readout}$ is
equal to 20\,s in the mode with binning $2 \times 2$, and to 90\,s with
binning $1 \times 1$. This is why, with the CCD one has to make exposures of
several minutes in each channel of the IFP. But here a problem arises of
taking into account the variations of atmospheric transparency and seeing in
individual CCD frames (photometric correction) because in reality only 1--2
scanning cycles can be performed. Because of this, with the advent of highly
sensitive photon counters superior to CCD at short exposures they start to be
extensively used for observations with the scanning IFP (Hernandez et al.,
2001). At the same time, the modification of the method of photometric
corrections of CCD frames (Section 3.4) allows one to make use of the
advantages of CCDs at long exposures.

\begin{figure*}
\centerline{
\psfig{figure=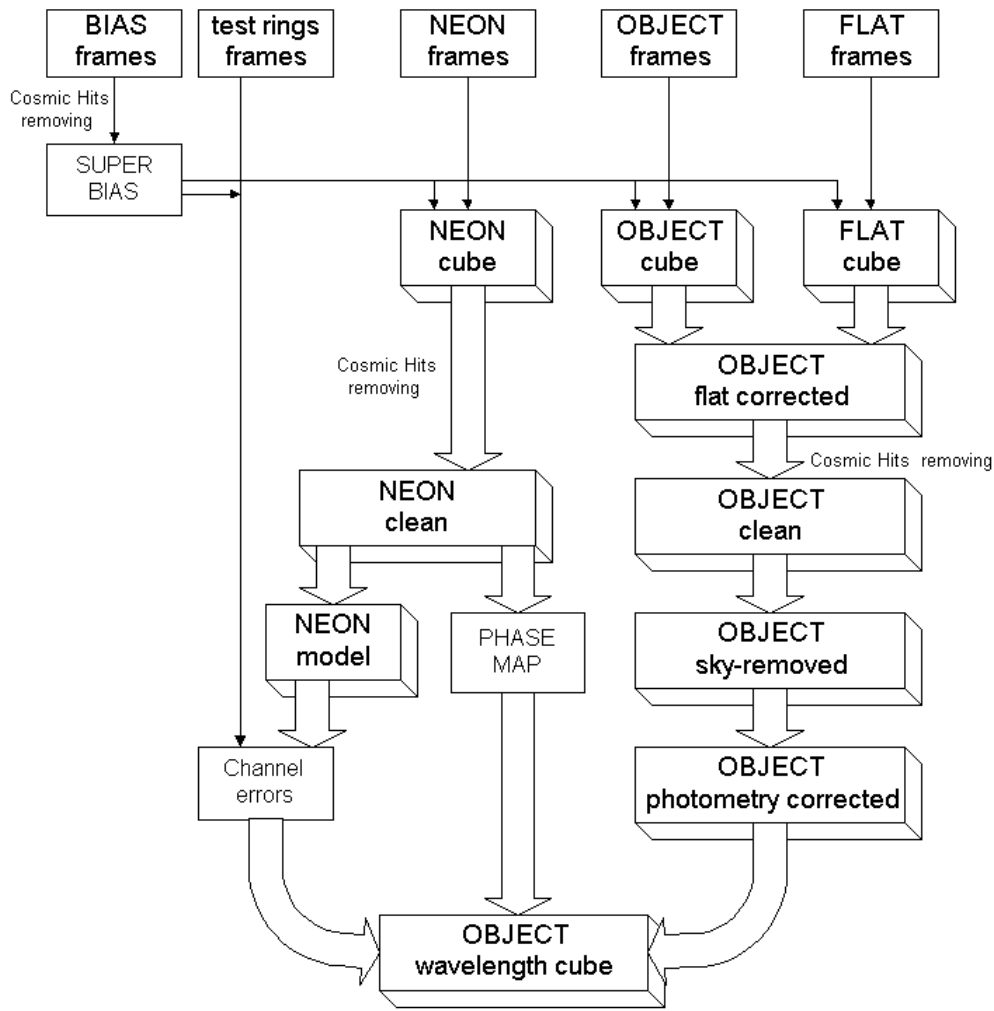,width=14cm}}
\caption{Block diagram of data reduction}
\label{scheme}
\end{figure*}

\section{Preliminary reduction and photometric correction}
\subsection{Raw data}
The following sets of CCD frames are original for reduction (see the block
diagram in Fig.\,4).
\begin{itemize}
\item \textbf{OBJECT} --- interferograms of the object under investigation. It is
recommended to observe first the images in the odd channels of the interferometer
(1, 3, 5...), then in all even (2, 4, 6...). This is done in order to avoid
systematic errors in evaluating the variations of atmospheric transparency
and seeing.
\item \textbf{NEON} --- images of interference rings from the emission line cut out
by a narrow-band filter from the spectrum of the built-in He-Ne-Ar lamp.
The calibration is generally performed before and after the observing night.
\item \textbf{FLAT} --- interferograms of uniform illumination of the ``flat field''
from the built-in continuous spectrum lamp, obtained with the same narrow-band
filter as was used in observing the object.
\item \textbf{TEST} --- images of the rings from the He-Ne-Ar lamp in individual
channels of the IFP which are obtained every 30--60 minutes during the
exposure of the object. They are used for monitoring the scanning accuracy of
the interferometer (the characteristics of could may be depend on time) and for
checking the shifts of the centers of the rings of the interference pattern,
which arise because of flexures of the apparatus. In the case of SCORPIO the
shift of the centre of the rings caused by flexures of the system reducer--CCD
does not exceed 0.5 pixel in the whole range of variations of zenith distances.
\end{itemize}

Bias frames (\textbf{BIAS}) and dark frames (\textbf{DARK}) ``standard'' for CCD observations
are also needed.\footnote{We won't speak further about taking account of
the dark frames, since in the CCD we use, it is about $0.1\bar{e}/\mbox{min}$,
i.e. it is insignificant at short exposures.}

\begin{figure*}
\psfig{figure=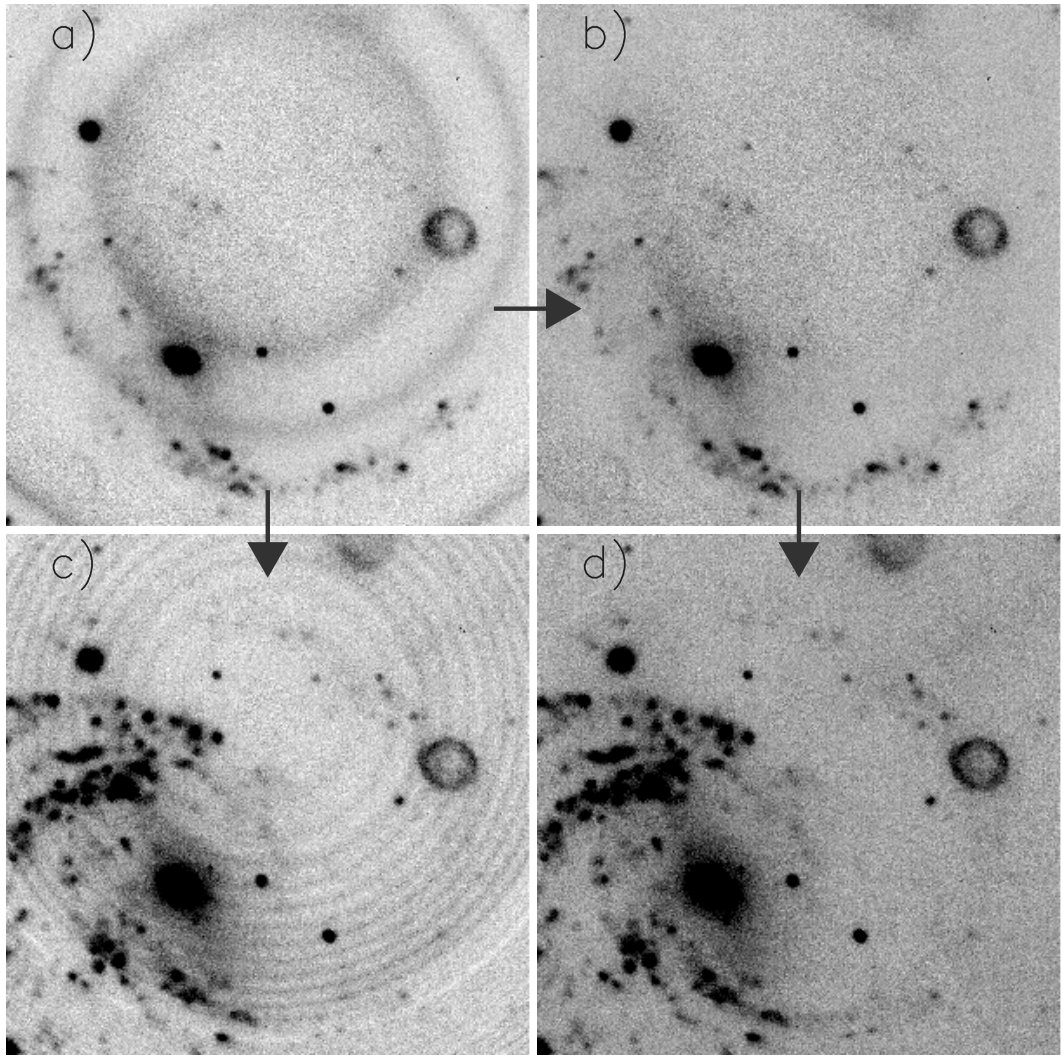,width=14cm}
\caption{Observations of the galaxy NGC\,6951 in the $H_{\alpha}$ line with
the  SCORPIO: a)  a fragment of the spectral channel with rings from the sky
lines; b)  the same frame after sky subtraction; c) the spectral channel
on the wavelength scale, which was obtained after the phase correction of the
cube with the sky (Method I); d) analogous channel after the phase correction
of the cube with the sky subtracted (Method II), the arrows indicate the
sequence of reduction in each of the method; e) the radial profile of the night
sky spectrum, the arrow indicates the dispersion direction.}
\label{sky}
\end{figure*}

\subsection{Building of data cubes and flat fielding}
The \textbf{BIAS} frames are averaged as in ordinary CCD reduction. The frame
SUPERBIAS obtained as a result of averaging is subtracted from all frames and
calibrations. Defective columns in the frames are masked. Further processing
is performed not of two-dimensional frames but of three-dimensional cubes (a
set of object's and calibration's frames). The removal of cosmic hits in the
cubes is done by a simple $\sigma$-filter: the counts in individual spectra
which differ from the mean by a threshold value (on the scale of standard
deviations) are replaced by the half-sum of the neighbouring ones.
Preliminarily, the cube \textbf{OBJECT} is normalized to the cube
\textbf{FLAT}, which makes it possible to take into account two effects at
once: firstly, the sensitivity variations of the CCD pixels and optics
throughput over the field of view (photometric ``flat field''), secondly, the
spectral modulation introduced by the narrow-band filter, which changes over
the detector field since the central wavelength of the interference filter
depends on the angle of beam incidence. Because of this, it is important that
the angle of convergence of the beam from the calibration lamp should be close
to the angle of convergence of the beam from the telescope. In other words,
the filling of the output pupil for the object and calibration must be equal.
This condition (the condition of telecentrism) is met in the SCORPIO by using
special optics in the calibration beam. The violation of the condition of
equivalency of the pupils leads, first of all, to shifting the spectra in
wavelength in the cube \textbf{FLAT}. A correct reduction of such data
(obtained with the old version of the optical reducer) is possible provided
that there are bright enough stars in the field, the spectrum of which can be
regarded as ``flat'' in a range of 10--20 \AA. In this case the shift in
$\lambda$ is defined from the cross-correlation of spectra of stars and flat
field (preliminarily corrected for the phase shift). After that a shift of
spectra by the required number of spectral channels was performed in the cube
\textbf{FLAT}, and the cube derived is used for normalizing \textbf{OBJECT}.
We will emphasize that in this case the normalization to \textbf{FLAT} should
be accomplished before the procedures of photometric and phase correction. The
residual modulation is corrected after the conversion of the cube to the
wavelength scale.

\subsection{Night sky line subtraction}
After the preliminary reduction described above the observed flux in each
channel can be represented in the form
\begin{equation}
\label{skyterms}
 I_{obs}(x,y)=(I_{obj}(x,y) \bullet PSF)\cdot Ext+I_{sky}(r),
\end{equation}
here $I_{obj}$ is the image of the object corrected for the atmosphere in the
given channel, $Ext$ is the atmospheric extinction, $I_{sky}$ is the flux
from the night sky modulated by the interferometer, $PSF$ is the point-spread
function.  The main problem is presented by: the night sky
brightness (emission spectrum, illumination from the Moon, etc.), the seeing
provided by $PSF$ and the atmospheric extinction $Ext$, which are
 dependent on the time
of observations. Apart from the smooth constant variations connected with the
zenith distance variations, random variations of all these quantities are also
of importance. In the universally accepted scheme of processing (Bland and
Tully, 1989; Boulesteix, 2000; Gordon et al., 2000), channel-by-channel
correction of the seeing and atmospheric transparency is performed first, then
follows the phase correction, while the sky line subtraction is accomplished
on the wavelength scale. Further, this method will be designated as Method 1.
The scheme is ideal when working with the photon counter, when the atmospheric
effect is averaged in each channel by multiple exposures. In observations with
CCD, such a technique is applicable when working in a spectral range free from
night sky emission lines. However, it is not infrequent that variations of
intensities of night sky lines lead to the fact that on the wavelength scale
the profile of the sky emission lines varies with radius $r$. In other words,
in the wavelength domain, artifacts (contrast rings)  appear in the channels
\textbf{OBJECT}. This imposes restrictions on the measurement accuracy of
velocities from weak emission lines of the object and causes systematic errors
in their estimations (Section 5.2).

For this reason, we proposed a method of sky subtraction in each image
prior to the phase correction (hereafter Method II). In the detector regions
free from the emission of the object the night sky lines are averaged by the
azimuthal angle in narrow rings of 0.5--1 pixel wide with the centre on the
optical axis of the IFP (Fig. 5). The derived radial profile (Fig. 5e) is
subtracted from the image (Fig. 5a,b). By repetition of this procedure
in all the channels one can correctly get rid of the effect of the second
term in (7) without introducing distortions in the wavelength domain.\footnote{When
this paper was submitted to the press, a paper of Jones et al. (2002) came out,
in which the authors used an analogous procedure for sky subtraction in
working with adjustable filters, which is a variety of the IFP operated
in small orders of interference.} Fig. 5 demonstrates clearly the advantage
of the procedure that we have adopted (sequence $a\rightarrow b\rightarrow d$)
as compared to Method 1 (sequence $a\rightarrow c$).

Clearly, the procedure of sky subtraction is sensitive to the accuracy
of determination of the centre of the rings (see Section 5.2) which may be
shifted from frame to frame. For automated search for an optimum position
of the centre of the rings from the sky, we apply a simple iterative procedure.
It is based on the minimization of the squares of deviations of counts in
the pixels from the mean profile in the detector region free from the object
emission lines. Fig. 6 shows the variation of the centre of the rings during
a 2-hour exposure, the zenith distance being varied from $z=15^\circ$ to
$z=45^\circ$. The scatter of points with respect to the main trend is most
likely to be caused by errors of finding the centre. They are usually
not larger than 0.05--0.2 pixels (depending on the intensity of sky lines),
which is sufficient for accurate sky subtraction. Fig. 6 also displays
relative variations of intensities of sky lines in observations of different
objects.

\begin{figure*}
\psfig{figure=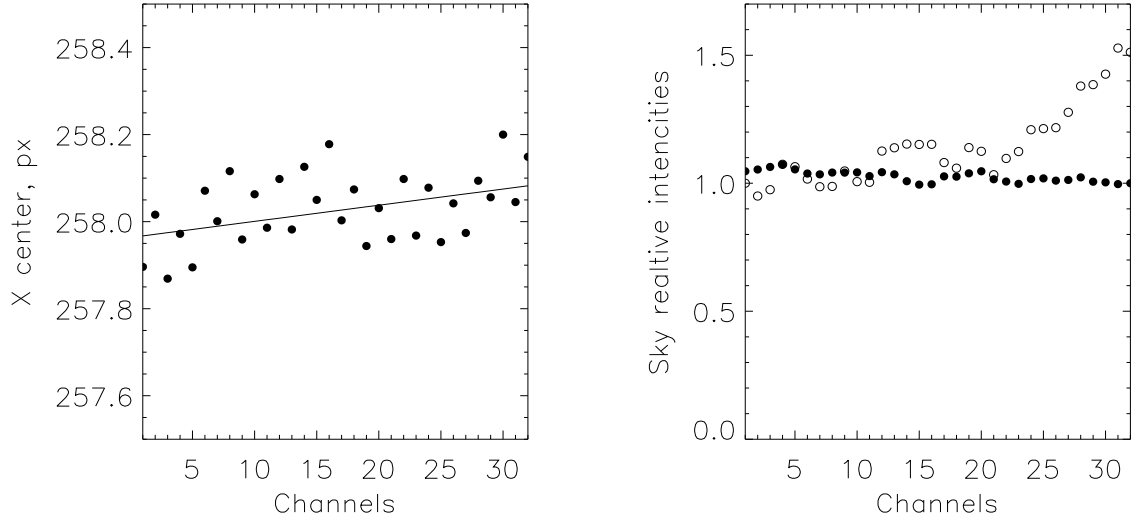,width=16cm}
\caption{Measurements of the position of the centre of the rings from the
sky in the object cube. (left panel). Measurements of the relative night sky
line intensities in observations on different nights (black dots and circles).}
\label{skyvar}
\end{figure*}

\subsection{Photometric correction}
Photometry of stars in the field of view makes it possible to evaluate the
contribution of atmospheric instability and guiding errors in each channel
\textbf{OBJECT}. With the aid of approximation by two-dimensional gaussians
relative shifts of the image centre, variations of the $FWHM$ and integral
flux are determined for each star. The variation of these parameters vrs. the
channel number in IFP observations with the SCORPIO found from 12 stars are
shown in Fig. 7. The frame's off-sets with respect to one another connected
with the telescope guiding errors do not exceed $0.2''-0.5''$ and are
corrected by the corresponding
 shifts of the frames (see Fig. 7). To correct the seeing, the frames are
convolved with two-dimensional gaussians, so that the resulting $FWHM$, the
same as in a frame with the worst images could be obtained. $FWHM$ variations
of several tens of per cent are thus diminished to 1--5\,\%, although, on
the whole, the spatial resolution in the cube becomes worse (Fig. 7).
Variations of atmospheric transparency are allowed for under the assumption
that the continuum in stars is flat in the narrow spectral range being
explored. In general, this is not always the case (especially around the
$H_{\alpha}$ absorption). However, if 8--10 stars located at different
$r$ are used, the mean flux variations correspond, as a rule, to the
extinction variations and do not depend on spectral features in stars.

\begin{figure*}
\psfig{figure=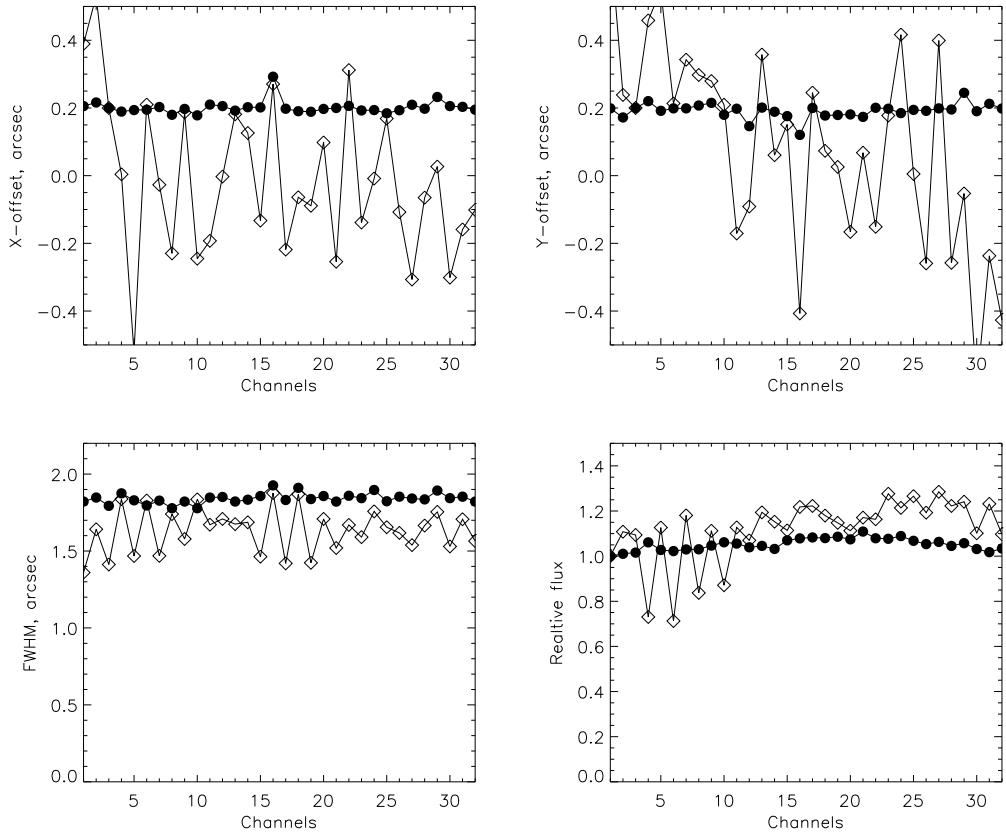,width=14cm}
\caption{Variations of the mean parameters of the field stars: relative shifts
of barycentres along $X$ and $Y$, the seeing, integral flux in relative units.
The open and filled symbols correspond to the observations of NGC\,6951
before and after the photometric correction.}
 \label{stars}
\end{figure*}

\begin{figure}
\centerline{
\psfig{figure=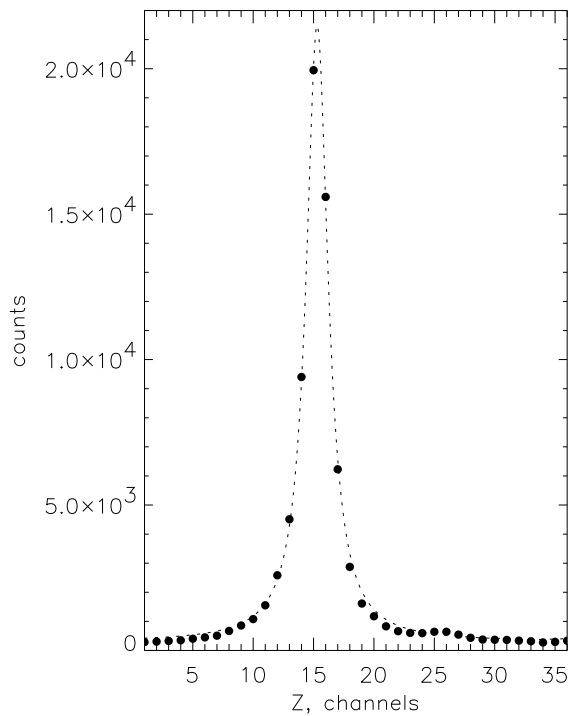,width=8cm}}
\caption{An example of the spectrum of the calibration line HeI$\lambda6678$
scanned with the interferometer Fp500 (dots) and its fitting by the Lorentz
profile.}
 \label{lore}
\end{figure}

\section{Wavelength scale calibration}
\subsection{Phase map}
According to (6), all the spectra in the observed cube are shifted linearly
(to an accuracy of a factor of $(r/f)^4$) by the value of the phase shift
with respect to a certain initial wavelength. For determination of the phase
shift the Lorentz profile was inscribed into the calibration line spectrum
in each point ($x,y$) of the cube \textbf{NEON}:
\begin{equation}
 I(x,y,z)=\frac{I_o(x,y)}{1+\left(\frac{2(z-p(x,y))}{w(x,y)}\right)^2},
\label{lorenc}
\end{equation}
which, according to Bland-Hawthorn (1995), is a much better approximation
of the hardware instrumental profile of the IFP as compared to the gaussian one.
Here $w(x,y)$ is the line width, while $p(x,y)$ is the phase shift sought for.
An example of the distribution $p$ over the field is displayed in Fig. 3.

\subsection{Channel-by-channel correction}
The method described above is used to find the phase shift assuming that the
quantities $A$ and $B$ that relates the channel number $z$ to the gap between
the plates (see Section 2.1) are constant in the course of scanning. There
are, however, a number of factors, such as temperature variations, errors and
failures in the electronics of control, which have an effect on the scanning
stability. For control of mutual shifts of the calibration channels for each
frame of the cube \textbf{NEON}, the best position with allowance made for the
shifts in all three coordinates inside the model cube $I(x,y,z)$ constructed
in accordance with (8) is sought by the least-squares method. For the IFP
operating steadily the scatter of scanning errors in $z$ inside one cube does
not exceed 1--3\,\% and does not affect the quality of data. Over the past 2
years of observations, channels with a scanning error of 10--20\,\%, which are
probably associated with noises in the electronics, have been present only two
times in the calibration cube. The above-described procedure revealed and
removed systematic differences of $\simeq10\,\%$ between the scanning step and
the correct value, which are caused by inaccurate selection of the constants
$A$ and $B$ in the observations of 2000.

\begin{figure*}
\psfig{figure=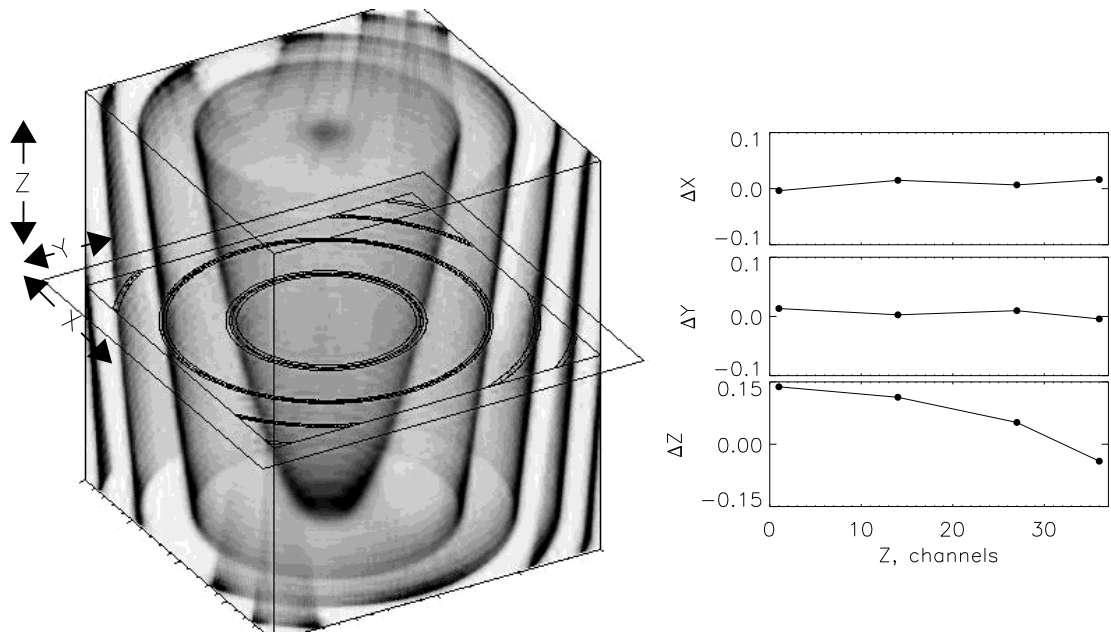,width=15cm}
\caption{A schematic view (left panel) illustrating search for the
optimum position of the image of the calibration rings inside the model
calibration cube. The gradations of the grey show the variations of intensity of the
model profile in the cube. Relative shifts of test acquisitions (in pixels)
relative to the calibration cube during a 2-hour exposure (right-hand panel).}
\label{model}
\end{figure*}
\begin{figure*}
\psfig{figure=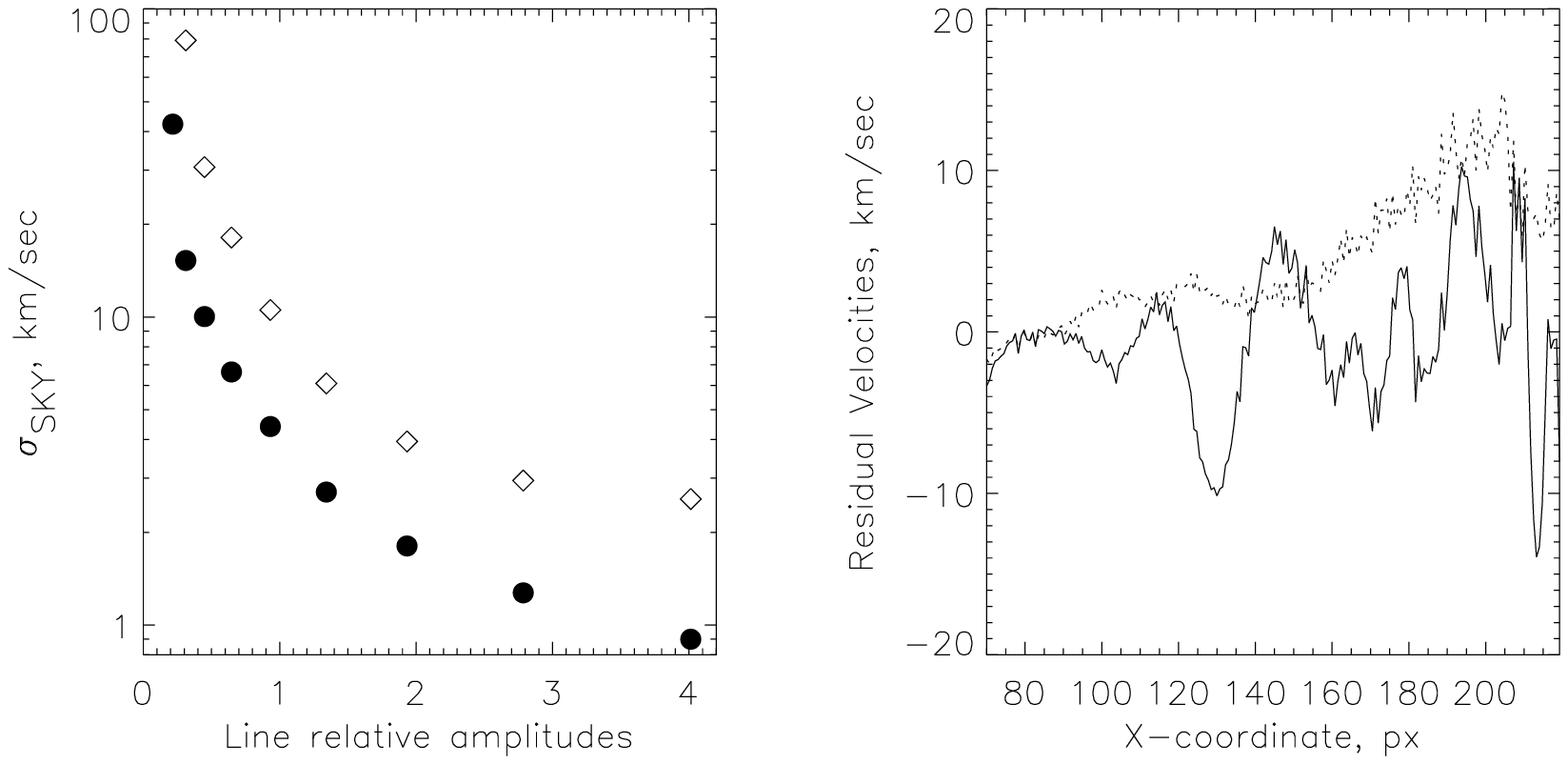,width=15cm}
\caption{Modeling of the errors. The relation of errors of velocity
measurement and the emission line intensities of the object (in units of
intensity of the strongest line in the sky) in the case of smooth intensity
variations of the sky lines and atmospheric extinction by 20\,\% during the
exposure. The open and filled symbols correspond to sky subtraction before and
after the phase correction (left-hand panel). An example of residual velocities
(measurements minus model) along the section across the velocity field. The
solid line represents random shifts of the centre of the rings with
$\sigma=0.5''$, the dotted line shows the smooth shift of the centre by
$\sigma=1''$ during the exposure (right-hand panel).}
\label{artifacts}
\end{figure*}

A search for relative shifts of the test images of the calibratin rings \textbf{TEST} is
fulfilled analogously. Fig. 9 shows the variations of the position of the
calibration ring centre during 2 hours of observations. The shift of the
centre of the rings is insignificant; one can see a smooth variation of the
scanning step by more than 15\,\% with respect to the calibration cube.
This is why, prior to the correction of the phase shift, an appropriate
correction in the $z$ coordinate is introduced into the object cube.

\section{Estimates of the accuracy of velocity measurements}
Here, for the sake of convenience in representation, some estimates will
be made in units of the spectral channel, which is $15-18\km$ for observations
in the $H_{\alpha}$ line with interferometer FP501 and $35-40\km$ for the
interferometer FP235. The error of measurement from the emission line
in the cube of the object may be represented as
\begin{equation}
\sigma_{Vel}=\sqrt{\sigma_{gaus}^2+\sigma_{\lambda}^2+\sigma_{reduct}^2 },
\end{equation}
where $\sigma_{gaus}$ is the error of determination of the line centre
by the gauss-approximation method defined by the S/N ratio in the
line; $\sigma_{\lambda}$ is the accuracy of reduction to the wavelength scale;
$\sigma_{reduct}$ are the errors appearing at the stage of reduction and
related, first of all, with provision for the atmospheric instability.

The contribution of $\sigma_{gaus}$ is small. We have modeled its relation
to the S/N ratio by the Monte-Carlo method. For typical line widths ($FWHM =
4-10$ channels) and Poisson noises, it turns out that the error of determination
of the centre $\sigma_{gaus}<0.05-0.08$ of the channel even at a $S/N>4-5$.

\subsection{Phase correction error}
An independent evaluation of the phase shift correction accuracy was performed
when measuring velocities of the night sky emission lines in the data cube
obtained with the EP501 polarimeter in the observations of M71 (Section 6.2).
The lines $H_\alpha\lambda 6562.82$\,\AA\, and $\mbox{OH}\lambda 6568.779$\,\AA\,
from the adjacent interference order were used.\footnote{Wavelengths are
 taken from
Osterbrock et al. (1996).} The mean velocity was $2.2\km$ with a point-to-point
dispersion of velocities of $2.0\km$, that is, the accuracy of construction of
the wavelength scale is about 1.0 channel, or $\simeq2\km$ for FP501 and
$\simeq4\km$ for FP235.

\subsection{Systematic errors}
Portions of images from frames with different $z$ are present at each radius
$r$ in the image of the spectral channel for a fixed $\lambda$. That is why,
any frame-to-frame inhomogeneity  (variations of the sky lines, seeing, etc.)
leads to the appearance of systematic distortions in the long-wave cube, which
is dependent on the radius. In turn, this causes systematic errors (artifacts)
in the obtained monochromatic images in the velocity field too. We present
these errors as consisting of several independent components:
\begin{equation}
\sigma_{reduct}^2= \sigma_{SKY}^2+\sigma_{SHIFT}^2+\sigma_{FLAT}^2 \label{art}.
\end{equation}
Here $\sigma_{SKY}$ is the error introduced by the variability of the
background and night sky lines, $\sigma_{SHIFT}$ is the guiding error, and
$\sigma_{FLAT}$ is the error of division by \textbf{FLAT}.

The contribution of different components in (10) is estimated with the aid of
modeling the process of reduction. Distributions of radial velocity and
brightness (disk + nucleus + HII regions) with the parameters typical of the
observed galaxies were specified in the sky plane. These data were used to
construct the cube in $\lambda$ domain, and having made a transformation, inverse
with (6), we obtained the ``observed'' cube in $z$ domain. The cube with
the rings from the night sky was constructed in a similar manner. By combining the
two cubes and introducing in them the required distortions (relative shift of
channels, transmission variation, additional noises, etc.), we constructed
the ``observed'' velocity fields, which we compared with initial velocities.
The modeling was performed for the version of observations with FP235.
The following results were derived.
\begin{itemize}
\item
$\sigma_{SKY}$ was evaluated for different variations of atmospheric extinction
and brightness of the night sky lines in the two methods of sky subtraction
(Section 3.3) --- Method I (in $\lambda$ domain) and Method II (in $z$ domain).
A characteristic relation between the velocity determination error and the line
intensity (in units of the brightest sky line $A_{sky}$) is displayed in Fig.
10. Even at 10--20 per cent variations of intensity of the sky lines and
extinction, Method II yields a velocity measurement error 2--5 times as low as
Method I, which is more essential for weak (with respect to the sky) lines of
the object. One may consider that $\sigma_{SKY} < 10\km$ (0.2 channel) even
for lines with an amplitude higher than $(0.3-0.5)A_{sky}$. The error becomes
insignificant (less than $2\km$) for lines with an amplitude $(1.5-2)
A_{sky}$. Naturally, $\sigma_{SKY}$ depends on the relative location of the
lines of the sky and object in the spectrum, but the accuracies given above
may be regarded to be typical. Fig. 10 illustrates the advantage of Method II
in the reduction of CCD observations of objects against the background of the
bright sky lines. If the intensity of the sky lines is insignificant, the two
methods give similar results.

When using Method II, it is important to precisely define the position of the
centre of the rings. The modeling has shown that $\sigma_{SKY}$ begins to
considerably increase if the error of determining the position of the centre
of the rings $\sigma_{CENTER}>0.3-0.5$ pixel (i.e. 0.05--0.08 minimum width of
the ring from the sky in the image). As was already noted in Section 3.3, this
accuracy is provided by the automated procedure of determination of the centre
of the rings.
\item
The errors of guiding and the flexures of the instrument lead to mutual shifts
between the images in the cube $z$ (see Fig. 7). The correction of this effect
from the position of the images of stars in the field causes mutual shift of
channels in the wavelength domain. As a result, an error in the determination
of velocities $\sigma_{SHIFT}$ arises. In contrast to $\sigma_{SKY}$, it is
only slightly dependent on the line intensity and is mainly defined by the
distance from the centre of the rings $r$. Fig. 10 shows the variations in
changes of the velocity in the cross-section of the image of the model galaxy
for two types of the channel shifts --- random from channel to channel
(guiding errors) and systematically increasing (result of instrumental
flexure). In the first case spurious rings arise, in the second case a
constant velocity gradient appears. Table 2 gives the mean amplitudes of
spurious velocities depending on the value of shifts of both types. The
systematic flexures in the SCORPIO do not exceed
 $0.2''-0.3''$
in the whole range of zenith distances, and the typical guiding errors
are of the order of $0.1''-0.2''$.
\begin{table}
\caption{Amplitudes of spurious velocities (in $\km$) for different channel
shifts}
\begin{center}
\begin{tabular}{ccc}
\hline
Shift ($''$) &\multicolumn{2}{c}{Type of shift}\\
     & random & systematic \\
0.05&    3.0    &         2.1 \\
0.10 &   5.5    &         2.2 \\
0.25 &   22     &         7.0 \\
0.50 &    25     &         16  \\
0.75&    32     &         23  \\
1.00 &    47     &         30  \\
\hline
\end{tabular}
\end{center}
\end{table}

\item
The transmission of the interference filter depends on the angle of ray
incidence. Because of this, if beams of light from the telescope and the
calibration beam do not coincide, the transmission curves of the filter in the
cube \textbf{FLAT} and \textbf{OBJECT} will then be different both in width
and in central wavelength. Using the photometry of the stars in the cube
\textbf{OBJECT}, this effect can be defined and corrected. The division by the
non-corrected cube \textbf{FLAT} will lead to distortions of the continuous
spectrum shape. Possible errors are determined by the relationship between the
filter width, 12--18\,\AA\, and by the free spectral range $\Delta\lambda$. It
can be easily shown that if the spectrum in the cube \textbf{FLAT} is shifted
by 2--3 channels or its $FWHM$ differs from \textbf{OBJECT} by 20\,\%,
perturbations will then arise with a dispersion of 0.15--0.2 and 0.05--0.08 of
continuum intensity for FP235 and FP501, respectively (at
$\lambda=6563$\,\AA\, with a filter $FWHM$ of 15\,\AA). When observing bright
emission lines with a weak continuum, this will cause appearance of an
insignificant error in measuring the velocity of $\sigma_{FLAT}$, however, for
the objects with a relatively strong continuum, the non-coincidence of the
beams from the object and calibration may result in appearance of spurious
lines in the spectrum at the background of the bright continuum.
\end{itemize}
\section{Results of data processing}
\begin{figure*}
\psfig{figure=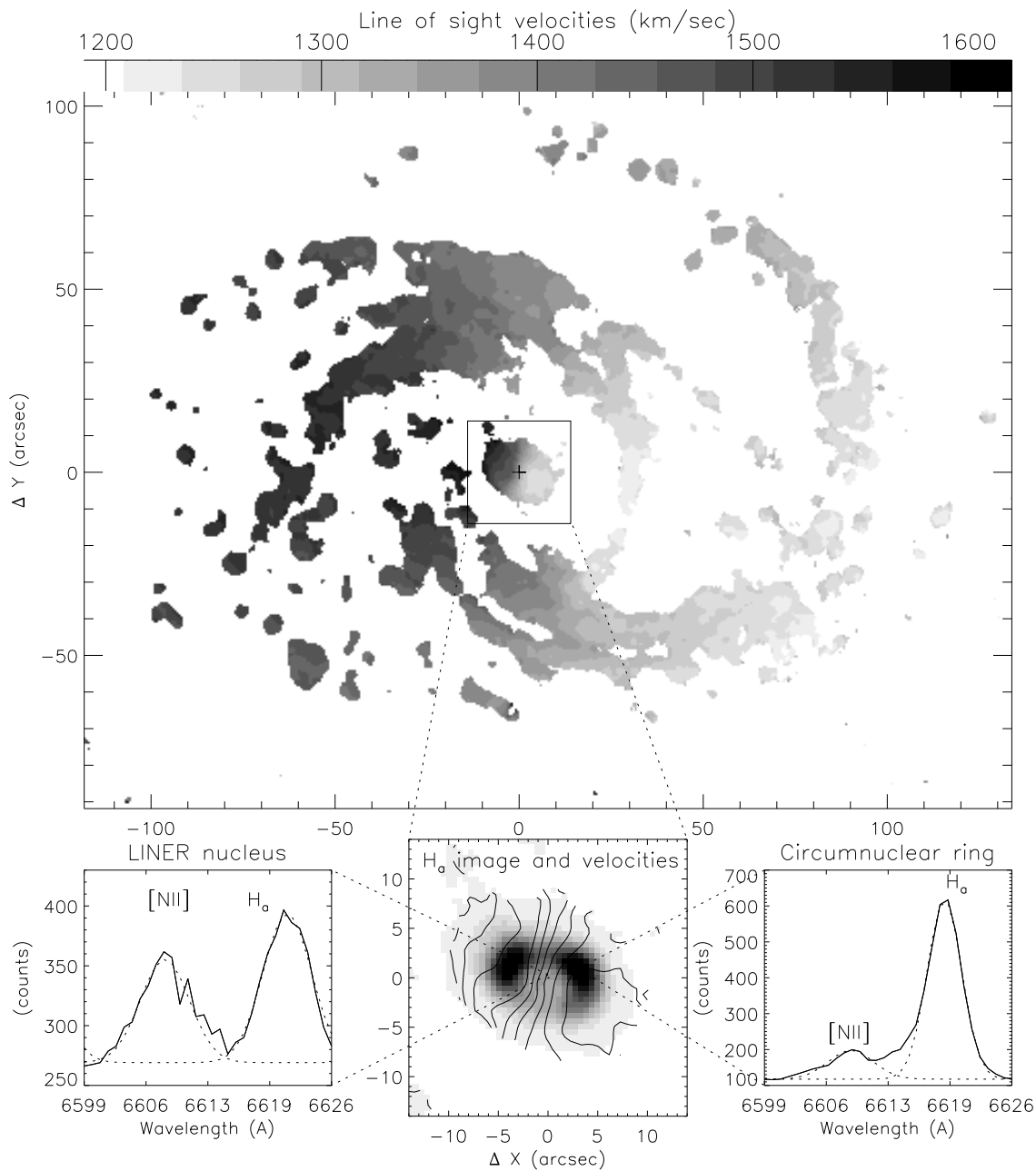,width=15cm}
\caption{Ionized gas in the galaxy  NGC~6951.
Top --- the velocity field in the  $H_\alpha$ line. The box cuts out the region
for which the image of the circumnuclear star formation ring in
$H_\alpha$ with lines of equal radial velocities is presented at the bottom.
The right- and left-hand panels give examples of Gauss approximation of the
profiles of the emission lines  $H_\alpha$ and
[NII]$\lambda6583$ in the nucleus and circumnuclear ring.}
\label{n6951}
\end{figure*}

\subsection{Emission object}
The galaxy NGC\,6951 was observed in the course of the first trial of the SCORPIO
 with the Fabry-Perot interferometer
23/24.IX.2000. The interferometer FP235 was employed. The scanning cycle
consisted of 32 frames with an exposure of 120 s for each. To increase the
readout rate a binning of $2\times 2$ was applied, the resulting pixel size is
$\sim0.56''$. A filter 16\,\AA\, wide centered on $\lambda_c=6596$\,\AA\,
separated a region around the $H_\alpha$ line of the galaxy. The variations of
the seeing and the atmospheric transparency before and after the photometric
correction are shown in Fig. 7, those of the sky line intensity are given in
Figs. 5 and 6. After the reduction to the wavelength scale, optimum filtration
of data was done, i.e. smoothing in the $z$ coordinate by a gaussian with a
$FWHM=1.5$ of the channel and in the ($x,y$) plane by a two-dimensional
gaussian $FWHM=2\times 2$ pixels. For this purpose, we used the package
ADHOC.\footnote{The package ADHOC was written by J. Boulesteix  (Marseille
Observatoire) and is available in Internet.} After smoothing, the spatial
resolution was about $2.7''$.

Apart from the $H_\alpha$ line, the line [NII]$\lambda6582$\,\AA\, is present
in the observed spectrum of the circumnuclear region. Using a fitting by gaussian
of the profiles of emission lines, we have constructed the velocity field
of ionized gas and a monochromatic image of the galaxy in each line. An
example of the velocity field in the $H_{\alpha}$ line is presented in Fig. 11.
In the same figure are shown examples of decomposition of the profiles of
emission lines in the LINER nucleus and the circumnuclear ring of star
formation. The velocities that we have measured are fully consistent with the data of
P\'erez et al. (2000) which they have obtained for this galaxy with three
positions of the spectrograph slit.

\subsection{Absorption object}
The central part of the globular cluster M71 was observed on 12/13.IX.2001
with the device SCORPIO and interferometer FP501 in the line $H_{\alpha}$ at
the request of N.N. Samus and A.S. Rastorguev. The scanning cycle consisted of
36 frames with an exposure of 120 s per frame; the same as in the preceding
case a binning of $2\times 2$ was applied. The seeing varied from $1.0''$ to
$1.7''$ during the observations. In each frame there are present monochromatic
images of stars at the wavelengths defined by their coordinates on the CCD
according to (1) (if one disregards the phase shift variations on the scale of
the image size). After the preliminary reduction and channel-by-channel
subtraction of the night sky lines, we made an automatic measurement of fluxes
from individual stars. Bright isolated stars were used to plot a mean
two-dimensional point-spread function (PSF) in each frame. Using the
cross-correlation method, the PSF was made coincident with the barycentre of
the star, the integral flux was defined by minimization via the least-squares
method. This simple method proved to be stable enough for tight stellar fields
and made it possible to do relative photometry of neighbouring stars with a
separation above $2.5''-3''$. For measuring velocities we used two different
methods. In the first variant, photometry of stars was performed in each
channel prior to the correction of the cube for the phase shift. The spectra
thus obtained $F_{\lambda}(z)$ were transformed to the wavelength scale  using
the phase shift value corresponding to the position of the barycentre of the
star on the phase map. In the second variant the channels in the cube of the
object were smoothed for reducing to one value of seeing as described in
Section 3.4, so that the average size of the images in the cube was
$FWHM=1.75''\pm 0.03''$. Next, the whole cube was corrected for the phase
shift and only then channel-by-channel photometry was done, which resulted in
the star spectrum $F_{\lambda}(\lambda)$. The profile of the absorption line
$H_{\alpha}$ in the spectra was fitted by a gaussian. Thus, velocities of more
than 700 stars, down to the 18th stellar magnitude in the $V$ band, were
measured in the observed field (see Fig. 12). The two methods of extraction of
spectra of stars yielded close results without notable systematic difference.
The velocity dispersion of stars of the cluster in the second method turned
out to be somewhat lower, which is likely to suggest its higher reliability. A
comparison with the published data on radial velocity of stars in this cluster
shows no systematic shift in observations with the IFP. The error of
individual velocity measurements of stars down to $V\sim17\fm 5$ estimated
from a comparison of the measured velocity dispersion of stars with the
literature data was $2-4\km$.

\begin{figure*}
\psfig{figure=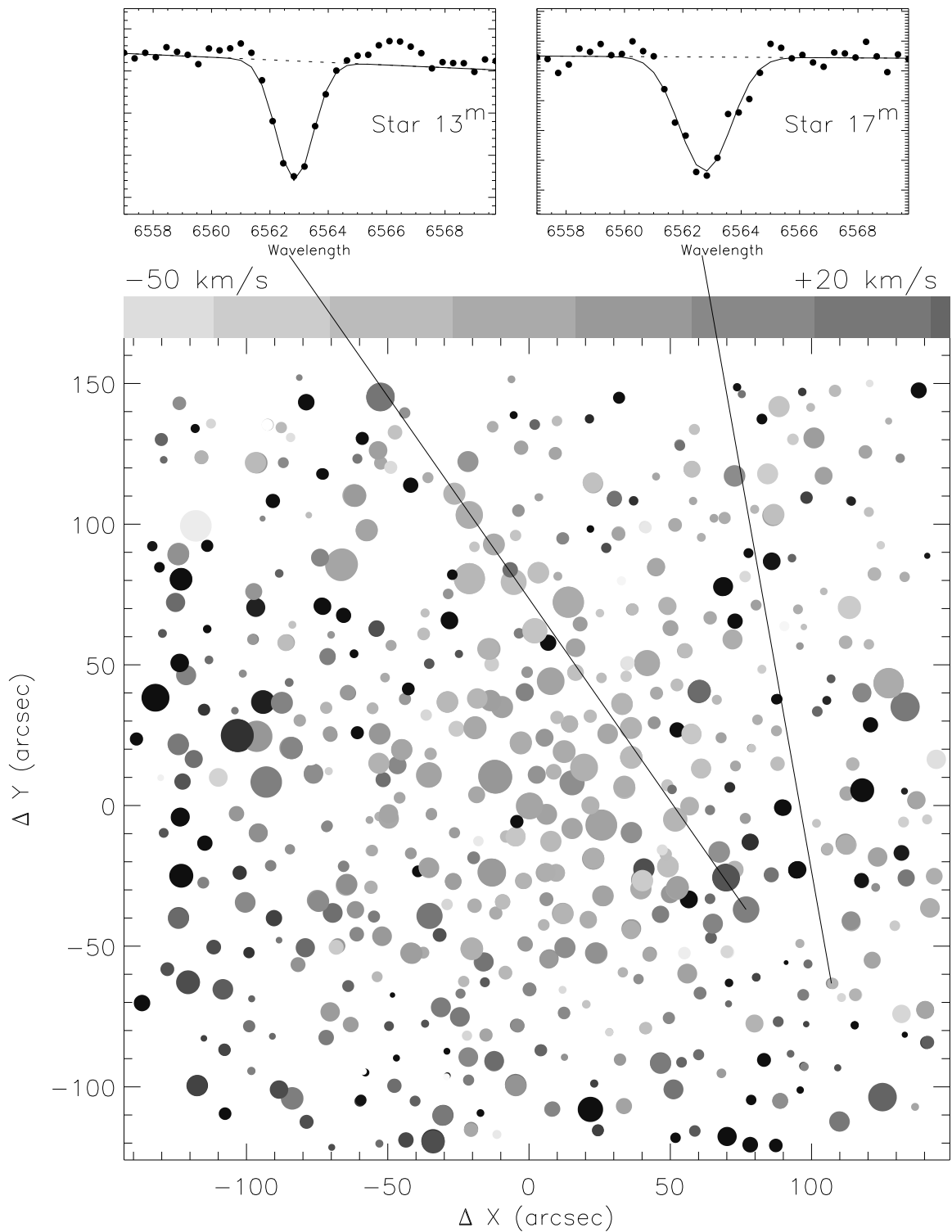,width=15cm}
\caption{Velocities of individual stars in the globular cluster M\,71.
The diameters of the circles are proportional to the stellar magnitude,
the colour --- to the radial velocity according to the scale at the top. Examples
are shown of  $H_\alpha$ absorption in the spectra of stars of  $V=13^m$ and
$V=17^m$ and their Gauss fitting.}
\label{m71}
\end{figure*}

\section{Conclusions}
The paper presents algorithms of reduction of observations of extended objects
with a scanning Fabry-Perot interferometer, which are oriented to a CCD as the
detector. The distinctive feature of such observations is that the radial
velocity measurement accuracy is often defined not by the formal value of the
S/N ratio, but by the presence of spurious rings in the wavelength domain.
Both the causes of their origin and the methods of their control at the stage
of data reduction are studied. All the above-described algorithms are
successfully applied to reduction of observations obtained with the optical
reducer SCORPIO of the 6m telescope prime focus.

\begin{acknowledgements}
The author thanks V.L. Afanasiev and T.A. Movsesian for useful comments and
also N.N. Samus and A.S. Rastorguev for permission to make use of the data
on M71 before their publication. The work was supported by the RFBR through
grants 01-02-16118, 01-02-17597 and by the Federal Programme ``Astronomy''
(Project 1.2.3.1).
\end{acknowledgements}

\end{document}